\documentclass[conference]{IEEEtran}

\usepackage{epsfig,makeidx,color,mathtools,bigints,graphicx,amsbsy,amsmath,amssymb,euscript,verbatim}

\newtheorem{remark}{Remark}
\newtheorem{example}{Example}
\newtheorem{thm}{Theorem}

\DeclareMathOperator*{\argmin}{arg\,min}

\DeclareMathOperator*{\diag}{diag}
\DeclareMathOperator*{\SINR}{SINR}

\newcommand{\R}{\mathbf{R}} 
\newcommand{\C}{\mathbf{C}} 

\begin{document}

\title{Hybrid Channel Pre-Inversion and Interference Alignment Strategies}

\author{
  \IEEEauthorblockN{David A. Karpuk}
  \IEEEauthorblockA{Dept. of Mathematics and Systems Analysis\\
    Aalto University\\
    P.O.\ Box 11100\\
    FI-00076 Aalto, Finland\\
    email: david.karpuk@aalto.fi} 
    \and
    \IEEEauthorblockN{Peter Moss}
    \IEEEauthorblockA{BBC Research \& Development\\
    South Lab, BBC Centre House \\
56 Wood Lane \\
London W12 7SB, England\\
email: peter.moss@rd.bbc.co.uk}
}

\maketitle

\IEEEpeerreviewmaketitle

\begin{abstract}
 In this paper we consider strategies for MIMO interference channels which combine the notions of interference alignment and channel pre-inversion.  Users collaborate to form data-sharing groups, enabling them to clear interference within a group, while interference alignment is employed to clear interference between groups.  To improve the capacity of our schemes at finite SNR, we propose that the groups of users invert their subchannel using a regularized Tikhonov inverse.  We provide a new sleeker derivation of the optimal Tikhonov parameter, and use random matrix theory to provide an explicit formula for the SINR as the size of the system increases, which we believe is a new result.  For every possible grouping of $K = 4$ users each with $N = 5$ antennas, we completely classify the degrees of freedom available to each user when using such hybrid schemes, and construct explicit interference alignment strategies which maximize the sum DoF.  Lastly, we provide simulation results which compute the ergodic capacity of such schemes.

\end{abstract}


\IEEEpeerreviewmaketitle

\section{Introduction}

Beginning with the seminal paper \cite{jafar} of Cadambe and Jafar, the promise of interference alignment to greatly increase capacity in the presence of interferers has made it a popular topic in recent years.  For a $K$-user interference channel, as the number of parallel links and the SNR approach infinity, coding strategies exist which guarantee all users $K/2$ degrees of freedom (DoF).  In the parlance of the interference alignment literature, ``everyone gets half the cake''.

The situation is quite different for constant channel coefficients without symbol extension.  In particular for a fully symmetric MIMO interference channel with $K$ users, each with $N$ antennas at the transmitter and receiver, and each user demanding $d$ DoF, a fundamental result by Bresler, Cartwright, and Tse \cite{bct} tells us that
\begin{equation}\label{funbound}
d\leq 2N/(K+1)
\end{equation}
which can place severe restrictions on the available degrees of freedom when one does not code over time.  Related results for MIMO interference channels have been established in, for example, \cite{annapureddy,razaviyayn,tresch,chenwei}.  Interference alignment strategies generally assume each user has access to only their own data.  However, during channel pre-inversion \cite{swindlehurst} all transmitters share all of their data, allowing them to pre-multiply the data vector by the inverse of the channel matrix.  In the symmetric MIMO interference channel, this strategy eliminates interference to give every user $d = N$ degrees of freedom.

This paper begins to address how the strategies of interference alignment and channel pre-inversion interact with each other.  If all of the $K$ users share their data then interference alignment is unnecessary.  However, if $K$ is large enough, such a data sharing scheme may become unrealistic, thus we consider situations where $K$ users are partitioned into smaller groups into which they share their data.  Each group uses a channel pre-inversion strategy to clear interference within its own group, and interference alignment strategies are employed to eliminate interference from other groups.  This problem bears some resemblance to an interference broadcast channel \cite{chenyang1,chenyang2}, but our channel model is fundamentally different.  The main contributions of this paper are the following:
\begin{itemize}
\item[$\bullet$] We propose that individual groups invert their subchannels using a Tikhonov inverse (i.e.\ MMSE precoding), and provide an apparently new derivation of the optimal Tikhonov parameter.  Our Theorem \ref{hammer} uses random matrix theory to provide an explicit, accurate estimate of the resulting SINR.  We essentially show that for $K >> 0$ and large SNR, we have $\text{SINR (dB)} \approx \text{SNR (dB)}/2$.  
\item[$\bullet$] For $K = 4$ and $N = 5$, we study every partition of the users into groups who share their data, and classify completely the increases in degrees of freedom.  The results of this section are not intended to generalize to larger values of $K$ and $N$; these values were selected because there are enough partitions of $K = 4$ users to provide a fertile testing ground for our ideas, and $N = 5$ antennas is the minimum necessary to achieve $d = 2$ degrees of freedom.  We restrict to cases in which every user has at least $2N/(K+1)$ degrees of freedom, so that every user benefits from data sharing.  Lastly, we plot empirical ergodic capacity curves for each strategy.
\end{itemize}



\section{Channel Pre-Inversion} \label{tikhonov}

Suppose we have an $N\times N$ MIMO channel, modeled by the familiar equation
\begin{equation} \label{groupchannel} 
Y = HAX + Z
\end{equation}
in which $X = [x_1,\ldots,x_N]^T$ with $x_i \in \C$ is the data vector, $A\in \C^{N\times N}$ is an encoding matrix, $Z$ is additive Gaussian noise, and the channel matrix is $H = (h_{ij})_{1\leq i,j\leq N}$ with $h_{ij}\in \C$.  Here we impose the energy constraint $\mathbf{E}(|x_i|^2) =1$ for all $i$, and we assume the entries $z_i$ of the noise vector $Z$ are assumed i.i.d.\ zero-mean Gaussian with variance $\sigma^2$ per complex dimension.  We assume that $H$ is known at the transmitter, but not necessarily at the receiver.  We further assume that the entries of $H$ are continuously distributed, so that, for example, $H^{-1}$ exists with probability $1$.

\subsection{Tikhonov Regularization}
Completely inverting the channel in (\ref{groupchannel}) requires us to set $A = H^{-1}$.  However, if one of the singular values $s$ of $H$ is close to zero, then the corresponding singular value $s^{-1}$ of $H^{-1}$ will cause the average energy per transmitter to be enormous.  We therefore consider for any matrix $B$ the \emph{Tikhonov regularization}, an approximate inverse defined by
\begin{equation}
B_\alpha := B^\dagger(\alpha I + BB^\dagger)^{-1}
\end{equation}
where $\alpha>0$ is some fixed constant.  One can show directly from the above definition that if
\begin{equation}\label{svd1}
B = U\Sigma V^\dagger,\quad \Sigma := \text{diag} (s_i(B))
\end{equation}
is a singular value decomposition of a $K\times K$ matrix $B$, then
\begin{equation}\label{svd2}
B_\alpha = V \Sigma_\alpha U^\dagger,\quad \Sigma_\alpha := \text{diag}(s_i(B)/(s_i(B)^2+\alpha))
\end{equation}
is a singular value decomposition of $B_\alpha$.  Hence for ill-conditioned $B$, the Tikhonov regularization $B_\alpha$ dampens the effect of badly-behaved singular values.

Now let $G := H/\sqrt{N}$ and $A := G_\alpha/\sqrt{N}$.  A simple computation shows that our channel equation becomes
\begin{equation}\label{newchannel}
\boxed{Y = GG_\alpha X + Z, \text{ for } G = H/\sqrt{N}.}
\end{equation}
This normalization has some advantages: the optimal $\alpha$ is independent of $N$, and the asymptotics of the SINR as $N\rightarrow \infty$ becomes easier to study using random matrix theory.

\subsection{The Optimal Tikhonov Parameter}

We now address the issue of choosing the optimal $\alpha$, which we consider to be the one which maximizes the signal-to-interference-plus-noise ratio, or SINR.  This question was of coursed addressed in \cite{swindlehurst}, however we present an apparently new and sleeker derivation of the optimal Tikhonov parameter.  

The above choice of normalized Tikhonov inverse reduces us to studying the SINR for the channel (\ref{newchannel}).  The transmitters need to rescale by the Frobenius norm of the encoding matrix $G_\alpha/\sqrt{N}$ before transmission to achieve unit average energy per user.  The ideal signal and interference powers will hence be scaled by the same constant.  Before rescaling, the ideal signal has expected power $N$, the expected noise power is $N\sigma^2$, and the interference power is $||GG_\alpha - I_N||^2_F$.  Rescaling the signal and interference powers by dividing by $\frac{1}{N}||G_\alpha||^2_F$ gives us
\begin{equation}\label{sinrdef}
\boxed{
\SINR = \frac{N}{||G_\alpha||^2_F\sigma^2 + ||GG_\alpha-I_N||^2_F}
}
\end{equation}
Note that the denominator of this expression is the Frobenius norm of the MSE matrix, thus maximizing the SINR is equivalent to minimizing the mean square error, which is a standard measure of system performance.

One can deduce the equality $||GG_\alpha - I_N||^2_F = ||G_\alpha G - I_N||^2_F$ by considering singular value decompositions as in (\ref{svd1}) and (\ref{svd2}) and using the invariance of the Frobenius norm under orthogonal transformation.  Choosing the resulting optimal Tikhonov parameter $\alpha$ to maximize the above SINR is therefore equivalent to solving the optimization problem
\begin{equation}
\alpha_{\text{opt}} =\argmin_\alpha \left(||G_\alpha||^2_F\sigma^2 + ||G_\alpha G  - I_N||^2_F\right)
\end{equation}
which is simply an MMSE optimization problem with the well-known solution
\begin{equation}
\boxed{\alpha_{\text{opt}} = \alpha_{\text{MMSE}} = \sigma^2.}
\end{equation}


\begin{remark}An easy computation shows that $GG_\alpha = HH_{N\alpha}$, hence our derivation of the optimal Tikhonov parameter results in the same as in \cite{swindlehurst}, since they conclude that the optimal Tikhonov inverse is $H_{N\sigma^2}$.  
\end{remark}

\subsection{Behavior of SINR as $N\rightarrow\infty$}

Suppose from now on that the entries of our channel matrix $H$ are i.i.d.\ zero-mean Gaussian with variance $1$ per complex dimension.   The following theorem and proof use random matrix theory to study the growth of the expression (\ref{sinrdef}), providing a way to compute the asymptotic SINR explicitly to within some error introduced by Jensen's Inequality.

\begin{thm}\label{hammer}
For an $N\times N$ MIMO system using the encoding matrix $G_\alpha/\sqrt{N}$ for a constant $\alpha>0$, we have
\begin{equation}
\boxed{
\lim_{N\rightarrow \infty} \mathbf{E}_G\left(\frac{1}{\SINR}\right) =
\frac{\alpha + (\alpha + 1)\sigma^2}{\sqrt{\alpha(\alpha+4)}} - \frac{\sigma^2}{2}
}
\end{equation}
\end{thm}

\emph{Proof:}  First we summarize the necessary ideas from random matrix theory, for which our main reference is \cite{haagthor}.  Let $f:[0,\infty)\rightarrow\R$ be a bounded, continuous function.  If $B$ is a $N\times N$ matrix such that $B^\dagger = B$ with necessarily real, positive eigenvalues $\lambda_i$, we define
\begin{equation}
\text{tr}_N(f(B)) := \frac{1}{N}\sum_{i = 1}^N f(\lambda_i)
\end{equation}
The key theorem we will use is Corollary 7.8 of \cite{haagthor}:
\begin{equation*}\label{bigtheorem}
\lim_{N\rightarrow\infty}\mathbf{E}_G(\text{tr}_N(f(G^\dagger G))) = \frac{1}{2\pi}\bigintssss_0^4 f(x)\frac{\sqrt{x(4-x)}}{x}dx
\end{equation*}
Let us rewrite $1/\SINR$ as
\begin{equation}\label{sinr_recip}
\frac{1}{\SINR} = \frac{1}{N}\left(||G_\alpha||^2_F + ||GG_\alpha - I_N||^2_F\right).
\end{equation}
We use the above theorem to evaluate the asymptotic expectation with respect to $G$ of this expression.  

Let $\lambda_i$ for $i = 1,\ldots,N$ be the eigenvalues of $G^\dagger G$.  It follows from plugging the expressions (\ref{svd1}) and (\ref{svd2}) for the singular value decompositions of $G$ and $G_\alpha$ into (\ref{sinr_recip}) that 
\begin{align}
\frac{1}{\SINR} 
& = \frac{1}{N}\left(\sum_{i = 1}^N \frac{\lambda_i}{(\lambda_i+\alpha)^2}\sigma^2 + \sum_{i = 1}^N\frac{\alpha^2}{(\lambda_i+\alpha)^2} \right) \\
& = \text{tr}_N(f(G^\dagger G))
\end{align}
where $f:[0,\infty)\rightarrow \R$ is the function
\begin{equation}
f(x) = \frac{x}{(x+\alpha)^2}\sigma^2 + \frac{\alpha^2}{(x+\alpha)^2}
\end{equation}
Using Mathematica to compute the relevant integral yields
\begin{align*}
\frac{1}{2\pi}\bigintssss_0^4 f(x)& \frac{\sqrt{x(4-x)}}{x}dx = \frac{\alpha + (\alpha + 1)\sigma^2}{\sqrt{\alpha(\alpha+4)}} - \frac{\sigma^2}{2}.
\end{align*}
as desired.  \hfill $\blacksquare$\\

\vspace{-.5cm}
\begin{figure}[h]
\centering
\includegraphics[width = .45\textwidth]{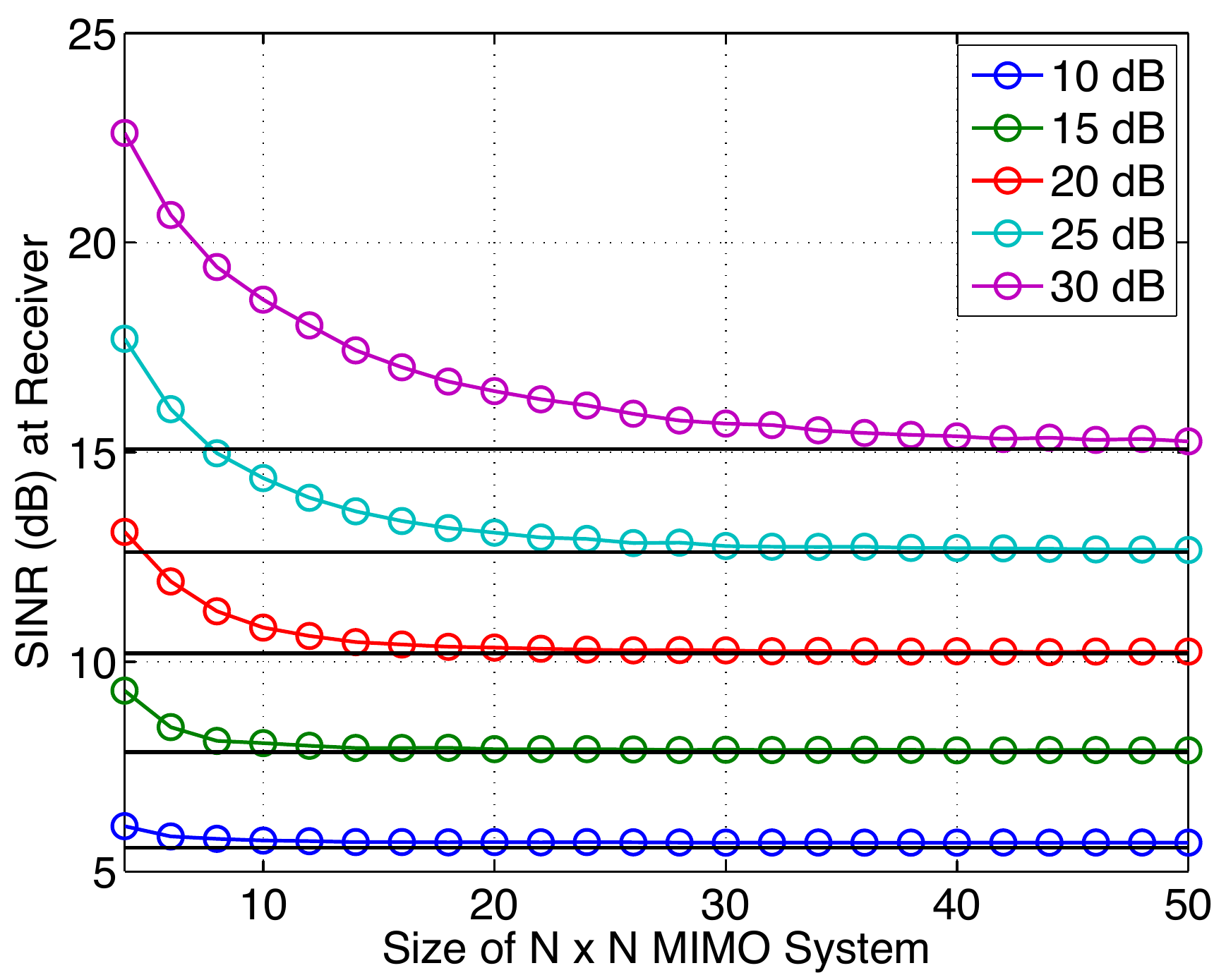}
\caption{Number of users employing the optimal $N\times N$ Tikhonov matrix $G_{\sigma^2}/\sqrt{N}$ versus SINR (dB), for various values of SNR (dB) $= 10\log_{10}(1/\sigma^2)$.  The horizontal black lines represent the values $\lim_{N\rightarrow\infty}1/\mathbf{E}_G(1/\SINR)$ as computed from Theorem \ref{hammer}, which serve as approximations for $\lim_{N\rightarrow \infty}\mathbf{E}_G(\SINR)$.}
\end{figure}

We can apply Jensen's Inequality to the convex function $1/x$, which tells us that
\begin{equation}\label{jensen}
\lim_{N\rightarrow \infty} \mathbf{E}_G(\SINR) \geq \lim_{N\rightarrow \infty} 1/\mathbf{E}_G(1/\SINR)
\end{equation}
thus Theorem \ref{hammer} allows us to predict the asymptotic SINR to within the error introduced by Jensen's Inequality.  As shown in Fig.\ 1., our simulations predict that this error is negligible, being $\approx 0.2$ dB at SNR $ = 10$ dB and less at higher SNR.  Notice that (\ref{jensen}) and Theorem \ref{hammer} also serve as a lower bound on the asymptotic SINR.

A simple computation shows that for sufficiently large SNR, one can use Theorem \ref{hammer} to deduce the following approximation for $\alpha = \alpha_{\text{opt}} = \sigma^2$:
\begin{equation}
\lim_{N\rightarrow \infty}\mathbf{E}_G(1/\SINR) \approx \sqrt{\sigma^2}.
\end{equation}
When combined with the estimate (\ref{jensen}) given by Jensen's Inequality, this provides us with the convenient expression for the SINR in terms of the SNR, for a large number of users $K$ and sufficiently large SNR:
\begin{equation}
\boxed{ \text{SINR (dB)} \approx \text{SNR (dB)}/2}
\end{equation}
One can also observe this experimentally from Fig.\ 1. \\

\section{Local Data Sharing}
\label{data_sharing}

We now suppose that we have a $K$-user MIMO interference channel, wherein each user has $N$ antennas.  We group the users together, such that within a group all of the users share their data and then can apply a Tikhonov pre-inversion scheme to their subchannel, as in the previous section.  

A \emph{partition} of a positive integer $K$ is a list $p(K) = (k_1,\ldots,k_m)$ of positive integers such that $\sum_{n = 1}^m k_n = K$.  Let us partition the $K$ users into $m$ groups, in which the $n^{th}$ group has size $k_n$.  Within each group the users share their encoded data $x_i$ at the transmit side, so that user $1$, for example, has access to $x_1,\ldots,x_{k_1}$.  The $n^{th}$ group then collaborates to encode their data with some $k_nN\times k_nN$ matrix $A_n$.  If we let $A$ be the block-diagonal matrix with the $A_n$ along the diagonal, we can rewrite the channel equation as
\begin{equation}
Y = \begin{bmatrix}
H_1 & * & * \\ * & \ddots & * \\ * & * & H_m
\end{bmatrix} 
\begin{bmatrix}
A_1 & 0 & 0 \\ 0 & \ddots & 0 \\ 0 & 0 & A_m
\end{bmatrix}
X + Z
\end{equation}
in which $H_n$ denotes the channel matrix for the $n^{th}$ group's subchannel.  The groups of users now set $A_n = (H_n)_\alpha$ for all $n = 1,\ldots,m$, resulting in an effective channel equation $Y = H'X + Z'$ where $H'$ has blocks of identity matrices of sizes $k_1N,\ldots,k_nN$ along the diagonal, and $Z'$ is a perturbation of $Z$ caused by the interference introduced by the failure of the Tikhonov inverse to be the exact inverse.  However, the interference introduced by this perturbation is minuscule when compared with the signal power and thus for the purposes of constructing the IA strategy, we treat it as noise and simply assume the diagonal blocks of the effective channel matrix are identity matrices.  The net effect of this process is that the users can eliminate all of the interference from the other members of their group.  We will study how this process affects the normalized sum DoF when we construct IA strategies with the new effective channel matrix.

Given a partition $p(K)$ of the users into groups who share their data, our goal is to
\[
\boxed{
\parbox{.4\textwidth}{
\begin{center}
compute $C_{p(K)} = \max\left\{\frac{1}{KN}\sum_{i = 1}^K d_i\right\}$ \\ subject to $d_i\geq 2N/(K+1)$ for all $i = 1,\ldots,K$
\end{center}
}
}
\]
where the max ranges over all possible interference alignment strategies.  The bound of \cite{bct} limits the above normalized sum DoF in the case of no data sharing, and we would like to improve on this bound if various combinations of users share their data.  We impose the constraint $d_i\geq 2N/(K+1)$ to limit us to cases in which no individual user is required to sacrifice any of the DoF they had in the case of no data sharing.


It is important to note that we do not assume any collaboration at the receive end.  For example, if users $1$ and $2$ collaborate, the signal from transmitter $2$ at receiver $1$ is still interference, whereas this was data at transmitter $1$.  In particular, breaking the users into groups of, for example, size $2$ does \emph{not} reduce our problem to the symmetric MIMO interference channel with $K/2$ users each with $2N$ antennas.

\section{Interference Alignment for MIMO Channels}

In this section we briefly summarize the necessary material concerning interference alignment for MIMO interference channels.   If user $i$ wants to transmit across $d_i \leq N$ dimensions (i.e.\ user $i$ has $d_i$ degrees of freedom), they use linear encoding to write $x_i = U_i \hat{x}_i$ for $\hat{x}_i \in \C^{d_i}$ a vector of information symbols (for example QAM symbols), and a full-rank $N\times d_i$ encoding matrix $U_i$. Receiver $i$ observes
\begin{equation}
y_i = H_{ii}U_i\hat{x}_i + \sum_{j\neq i} H_{ij}U_j\hat{x}_j + z_i
\end{equation}
Let us define the \emph{signal space} and \emph{interference space} at receiver $i$ to be, respectively,
\begin{equation}
\mathcal{S}_i := \text{colspan}(H_{ii}U_i), \ \mathcal{I}_i := \sum_{j\neq i}\text{colspan}(H_{ij}U_j).
\end{equation}
If $V_i$ denotes projection onto $(\mathcal{I}_i)^\perp$, receiver $i$ computes
\begin{equation}
V_iy_i = V_iH_{ii}U_i\hat{x}_i + V_iz_i
\end{equation} and can then reliably recover the desired signal $\hat{x}_i$ as long as $\dim(V_i \mathcal{S}_i)= d_i$.  With probability $1$, we have $\dim \mathcal{S}_i = \dim(V_i\mathcal{S}_i) = d_i$, provided
\begin{equation}\label{IAconstraint}
d_i + \dim \mathcal{I}_i \leq N.
\end{equation}
The goal of interference alignment is to choose matrices $U_i$ of rank $d_i$ for all $i=1,\ldots,K$ to
\begin{equation}
\boxed{
\parbox{.36\textwidth}{
\begin{center}
maximize $C = \frac{1}{KN}\sum_{i = 1}^K d_i$ subject to (\ref{IAconstraint}).
\end{center}
}
}
\end{equation}
We refer to $C$ as the \emph{normalized sum DoF}; multiplication by $\frac{1}{KN}$ makes meaningful comparison possible over different numbers of users and antennas.  We will call a choice of $U_1,\ldots,U_K$ satisfying (\ref{IAconstraint}) an \emph{interference alignment strategy}.

Existence results for IA strategies determine the feasibility of interference alignment and possible degrees of freedom, such as  \cite{jafar_IA_MIMO_feas,ghasemi_IA_MIMO}.   For our purposes these results have essentially been subsumed by the bound (\ref{funbound}) of \cite{bct}.  In addition to these fundamental limits, explicit IA strategies are often constructed using numerical optimization, as in \cite{heath_IA_algs, heath_IA_grassmannian}.  

\section{Data Sharing for $K = 4$ Users}

The bound of \cite{bct} gives us $d\leq 2N/5$, so let us consider the case $d = 2$, $N = 5$ as a base case.  Note that for all $i$ and $j$ we have $\max_{i\neq j} d_i\leq \dim \mathcal{I}_j$.  Hence by (\ref{IAconstraint}) we must have $d_j + \max_{i\neq j} d_i \leq N$, from which it follows that $\max_i d_i \leq 3$.  Thus our constraints force $d_i = 2$ or $3$ for all $i$.  Our results, which are derived in the following subsections, are summarized here:  
\vspace{-.5cm}
\begin{table}[ht]
\centering
\caption{Partitioning $K = 4$ Users with $N = 5$ Antennas into Groups}
\begin{tabular}{|c|c|c|c|}
\hline
$p(4) = $ sizes of groups & $(d_1,\ldots,d_4)$ & $C_{p(K)}$\\
\hline
$(1,1,1,1)$ & $(2,2,2,2)$ & $2/5$ \\
\hline
$(2,1,1)$ & $(2,2,2,2)$ & $2/5$ \\
\hline
$(2,2)$ & $(3,3,2,2)$ & $1/2$ \\
\hline
$(3,1)$ & $(3,3,3,2)$ & $11/20$ \\
\hline
$(4)$ & $(5,5,5,5)$ & $1$ \\
\hline
\end{tabular}
\end{table}
\vspace{-.3cm}

We should mention that our results contain the implicit assumption that the columns of the matrix $U_j$ are always linearly independent.  This can be justified using arguments similar to those in \cite{chenwei}, but should be intuitively clear as the columns of the $U_j$ are often distinct eigenvectors of random matrices.

\subsection{The Partition $p(4) = (1,1,1,1)$}  We start with the partition $p(4) = (1,1,1,1)$ and outline an explicit interference alignment strategy achieving the upper bound (\ref{funbound}), similar to that in \cite{tresch}.  We write $U_j = [U_{j}^{(1)}\ U_j^{(2)}]$ where $B^{(i)}$ denotes the $i^{th}$ column of a matrix $B$, and $U_j^{(1)}, U_j^{(2)}\in \C^{5\times 1}$.  To create the necessary $N-d=3$-dimensional interference space $\mathcal{I}_i$, we construct bases $\mathcal{B}_i$ of $\mathcal{I}_i$ of size $3$.  Explicitly, we choose
\begin{align*}
\mathcal{B}_1 &= \{ H_{12}U_2^{(1)}, H_{12}U_2^{(2)}, H_{13}U_3^{(1)}\} \\
\mathcal{B}_2 &= \{ H_{23}U_3^{(1)}, H_{23}U_3^{(2)}, H_{24}U_4^{(1)}\} \\
\mathcal{B}_3 &= \{ H_{34}U_4^{(1)}, H_{34}U_4^{(2)}, H_{31}U_1^{(1)}\} \\
\mathcal{B}_4 &= \{ H_{41}U_1^{(1)}, H_{41}U_1^{(2)}, H_{42}U_2^{(1)}\}
\end{align*}
The signal $U_1^{(2)}$, for example, interferes at receivers $2$ and $3$, thus $H_{21}U_{1}^{(2)} \in \mathcal{I}_2$ and $H_{31}U_1^{(2)} \in \mathcal{I}_3$ which gives a non-trivial linear relationship between the basis vectors of $\mathcal{B}_2$ and $\mathcal{B}_3$.   We continue in this manner, expressing each $U_j^{(i)}$ in two different ways, until we have 8 independent non-trivial linear combinations of the above basis vectors.  We solve the resulting $8\times 8$ linear system by first solving for $U_{4}^{(2)}$, substituting the result into the remaining equations, and continuing until we are left with an equation $AU_{1}^{(1)} = \lambda U_1^{(1)}$ for some generically invertible $A$ and some non-zero $\lambda$.  We finish by picking $U_1^{(1)}$ to be an eigenvector of $A$ with $\lambda$ the corresponding eigenvalue, and back-substitute to find the remaining $U_j^{(i)}$.  

\subsection{The Partition $p(4) = (2,1,1)$}

After the groups pre-invert their subchannels, we reduce the channel matrix to the form
\begin{equation}
H = \begin{bmatrix}
I_{N} & 0 & H_{13} & H_{14} \\
0 & I_N & H_{23} & H_{24} \\
 H_{31} & H_{32} & I_N & H_{34} \\ 
 H_{41} & H_{42} & H_{43} & I_N
\end{bmatrix}
\end{equation}
By symmetry, we can assume that $d_1 \geq d_2$ and that $d_3 \geq d_4$.  If $d_1 = 3$ then $\dim \mathcal{I}_i\geq 3$ for $i = 3,4$, which forces $d_3 = d_4 = 2$.  Suppose that the tuple $(d_1,d_2,d_3,d_4) = (3,2,2,2)$ were achievable.  Without loss of generality we can choose the following bases for the interference spaces:
\begin{align*}
\mathcal{B}_1 &= \{ H_{13}U_3^{(1)}, H_{13}U_3^{(2)}\} \\
\mathcal{B}_2 &= \{H_{24}U_4^{(1)}, H_{23}U_3^{(1)}, H_{23}U_3^{(2)} \} \\
\mathcal{B}_3 &= \{ H_{31}U_1^{(1)}, H_{31}U_1^{(2)}, H_{31}U_1^{(3)} \} \\
\mathcal{B}_4 &= \{ H_{41}U_1^{(1)}, H_{41}U_1^{(2)}, H_{41}U_1^{(3)} \}
\end{align*}

Successfully aligning the interference at the first two receivers would give us the following three equations:
\begin{align*}
H_{14}U_4^{(1)} &= a_1H_{13}U_3^{(1)} + a_2H_{13}U_3^{(2)} \\
H_{14}U_4^{(2)} &= b_1H_{13}U_3^{(1)} + b_2H_{13}U_3^{(2)} \\
H_{24}U_4^{(2)} &= c_1H_{23}U_3^{(1)} + c_2H_{23}U_3^{(2)} + c_3H_{24}U_4^{(1)}
\end{align*}
solving the above for $U_4^{(1)}$, $U_4^{(2)}$ and $U_3^{(2)}$ gives us expressions of the form
\begin{equation*}
U_3^{(2)} = G_1U_3^{(1)},\ U_4^{(1)} = G_2U_3^{(1)},\ U_4^{(2)} = G_3U_3^{(1)}
\end{equation*}
At the third and fourth receivers we express all of the incoming interference in terms of the bases $\mathcal{B}_3$ and $\mathcal{B}_4$, respectively, and use the above expressions to arrive at the equations
\begin{align*}
U_2^{(1)} &= \sum_{i = 1}^3d_iH_{32}^{-1}H_{31}U_1^{(i)} = \sum_{i = 1}^3 e_iH_{42}^{-1}H_{41}U_1^{(i)}\\
U_2^{(2)} &= \sum_{i = 1}^3f_iH_{32}^{-1}H_{31}U_1^{(i)} = \sum_{i = 1}^3 g_iH_{42}^{-1}H_{41}U_1^{(i)} \\
U_3^{(1)} &= \sum_{i = 1}^3h_iG_2^{-1}H_{34}^{-1}H_{31}U_1^{(i)} = \sum_{i = 1}^3 k_iH_{43}^{-1}H_{41}U_1^{(i)}\\
U_3^{(1)} &= \sum_{i = 1}^3m_iG_3^{-1}H_{34}^{-1}H_{31}U_1^{(i)} = \sum_{i = 1}^3n_iG_1^{-1}H_{43}^{-1}H_{41}U_1^{(i)}
\end{align*}
Rearranging gives us four non-trivial equations of the form $\sum_{i = 1}^3a_iA_iU_1^{(i)} = 0$ which are all independent with probability $1$.  Such a system has no solution, contradicting our original assumption that $d_1 = 3$.  A similar argument shows that we cannot have $d_3 = 3$, and we conclude that $C_{(2,1,1)} = 2/5$.

However, there is some benefit to users $1$ and $2$ inverting their subchannel, in that the interference alignment strategy is easier to construct.  Explicitly, we first solve the two independent alignment chains
\begin{align*}
\alpha_2H_{42}U_2^{(1)} &= H_{43}U_3^{(1)}\ \ & \beta_2H_{42}U_2^{(2)} = H_{43}U_3^{(2)} \\
\alpha_3H_{23}U_3^{(1)} &= H_{24}U_4^{(1)}\ \ & \beta_3H_{13}U_3^{(2)} = H_{14}U_4^{(2)} \\
\alpha_4H_{34}U_4^{(1)} &= H_{32}U_2^{(1)}\ \ & \beta_4H_{34}U_4^{(2)} = H_{32}U_2^{(2)}
\end{align*}
which aligns the interference at receivers $1$ and $2$.  We now need only to choose $U_1$ such that the two conditions
\begin{align}
\dim \mathcal{I}_3 &= \dim \text{colspan}(H_{31}U_1,H_{32}U_2) = 3 \\
\dim \mathcal{I}_4 &= \dim \text{colspan}(H_{41}U_1,H_{42}U_2) = 3
\end{align}
hold, which is easily done.  One can choose, for example, $U_1^{(1)} = H_{31}^{-1}H_{32}U_2^{(1)}$ and $U_1^{(2)} = H_{41}^{-1}H_{42}U_2^{(2)}$.

\subsection{The Partition $p(4) = (2,2)$}\label{example}

The two groups separately invert their subchannels to obtain 
\begin{equation}
H = \begin{bmatrix}
I_{2N} & * \\
* & I_{2N}
\end{bmatrix}
\end{equation}
We are free to assume that $d_1 \geq d_2$ and that $d_3 \geq d_4$.  We then align the interference at receivers $1$ and $2$ so that 
\begin{equation}
H_{14}U_4 \prec H_{13}U_3,\quad  H_{24}U_4 \prec H_{23}U_3
\end{equation}
and similarly at receivers $3$ and $4$, where $A\prec B$ means $\text{cols}(A)\subset \text{colspan}(B)$.  This gives the lone constraint equation $d_1 + d_3 = 5$.  It is easy to see an optimal choice is now given by $d_1 = d_2 = 3$, and $d_3 = d_4 = 2$, resulting in $C_{(2,2)} = 1/2$.

\subsection{The Partition $p(4) = (3,1)$}

Subchannel pre-inversion gives the effective channel matrix
\begin{equation}
H = \begin{bmatrix}
I_{3N} & * \\
* & I_N
\end{bmatrix}
\end{equation}
Suppose without loss of generality that $d_1 \geq d_2 \geq d_3$.  It is easy to align the interference at receiver $4$ by picking any $U_2$ and $U_3$ satisfying 
\begin{equation}
U_2 \prec H_{42}^{-1}H_{41}U_1 \text{ and } U_3\prec H_{43}^{-1}H_{42}U_2
\end{equation}
Hence we are free to pick $d_1 = d_2 = d_3 = 5-d_4$ without restriction.  The maximum of $C_{(3,1)} = (3(5-d_4) + d_4)/20$ is achieved when $d_4$ is minimized, thus $C_{(3,1)}$ obtains its maximum value of $11/20$ when $d_4 = 2$ and $d_1 = d_2 = d_3 = 3$.

\section{Simulation Results for $K = 4$ and $N = 5$}

In this section we study the capacity of the above hybrid channel pre-inversion and interference alignment schemes for $K = 4$ users each with $N = 5$ antennas, to empirically demonstrate the benefits of using the Tikhonov regularization in concert with an interference alignment strategy.   Let us first study the SINR per user when employing one of the above strategies.  Suppose that we have partitioned $K$ users into groups of sizes $(k_1,\ldots,k_m)$, and for simplicity let us write $k = k_1$ for the size of the first group.  The channel equation for the first group reads
\begin{equation}
Vy= VG_1G_{1\alpha} U\hat{x} + Vz
\end{equation}
where $G_1 = H_1/\sqrt{kN}$ is the normalized subchannel matrix for this group, $G_{1\alpha}$ it its Tikhonov inverse, $U = \diag(U_1,\ldots,U_k)$ is the block diagonal matrix whose blocks are the encoding matrices for the IA scheme, and $V = \diag(V_1,\ldots,V_k)$ is the block diagonal matrix whose blocks are the projection matrices for the IA scheme, which have eliminated the interference from the other $m-1$ groups.  Here $\hat{x} = [\hat{x}_1,\ldots,\hat{x}_k]^t$ with $\hat{x}_i\in \C^d$, $y = [y_1,\ldots, y_k]^t$ with $y_i\in \C^N$, and $z = [z_1,\ldots,z_k]^t$ with $z_i\in \C^N$.

Denoting by $Q$ the MSE matrix $G_1G_{1\alpha} - I_{kN}$, we rewrite the above as
\begin{equation}
Vy = VU\hat{x} + VQU\hat{x} + Vz
\end{equation}
and thus user $i$'s instantaneous SINR is given by
\begin{equation}\label{SINR_IA_def}
\text{SINR}_i = \frac{||V_iU_i||^2_F/E_i}{||V_i||^2_F\sigma^2 + I_i/E_i},
\end{equation}
\begin{equation}
I_i = \sum_{j = (i-1)d+1}^{id}||(VQU)_{(j)}||^2_F,\ \ 
E_i = \frac{1}{k}\left(\frac{||G_{1\alpha}||^2_F}{kN}\right)
\end{equation}
and where $(VQU)_{(j)}$ denotes the $j^{th}$ row of the matrix $VQU$.  Here $I_i$ is the interference power at receiver $i$, and $E_i$ is the expected power expended by one user in a $k$-user group of size $kN$ when using Tikhonov pre-inversion, which the users must rescale by to maintain the power constraint.

We define $\SINR_i$ in the obvious way for members of the other groups, and after computing the expectation of each $\SINR_i$ over a number of channel matrices, we can approximate user $i$'s normalized ergodic capacity, and the capacity of the whole system, by
\begin{equation}
R_i = \frac{d_i}{KN}\log_2(1+\text{SINR}_i),\ \ R = \sum_{i = 1}^K R_i
\end{equation}
which we plot in Fig.\ 2 as a function of SNR.
\begin{figure}[h]
\centering
\includegraphics[width = .47\textwidth]{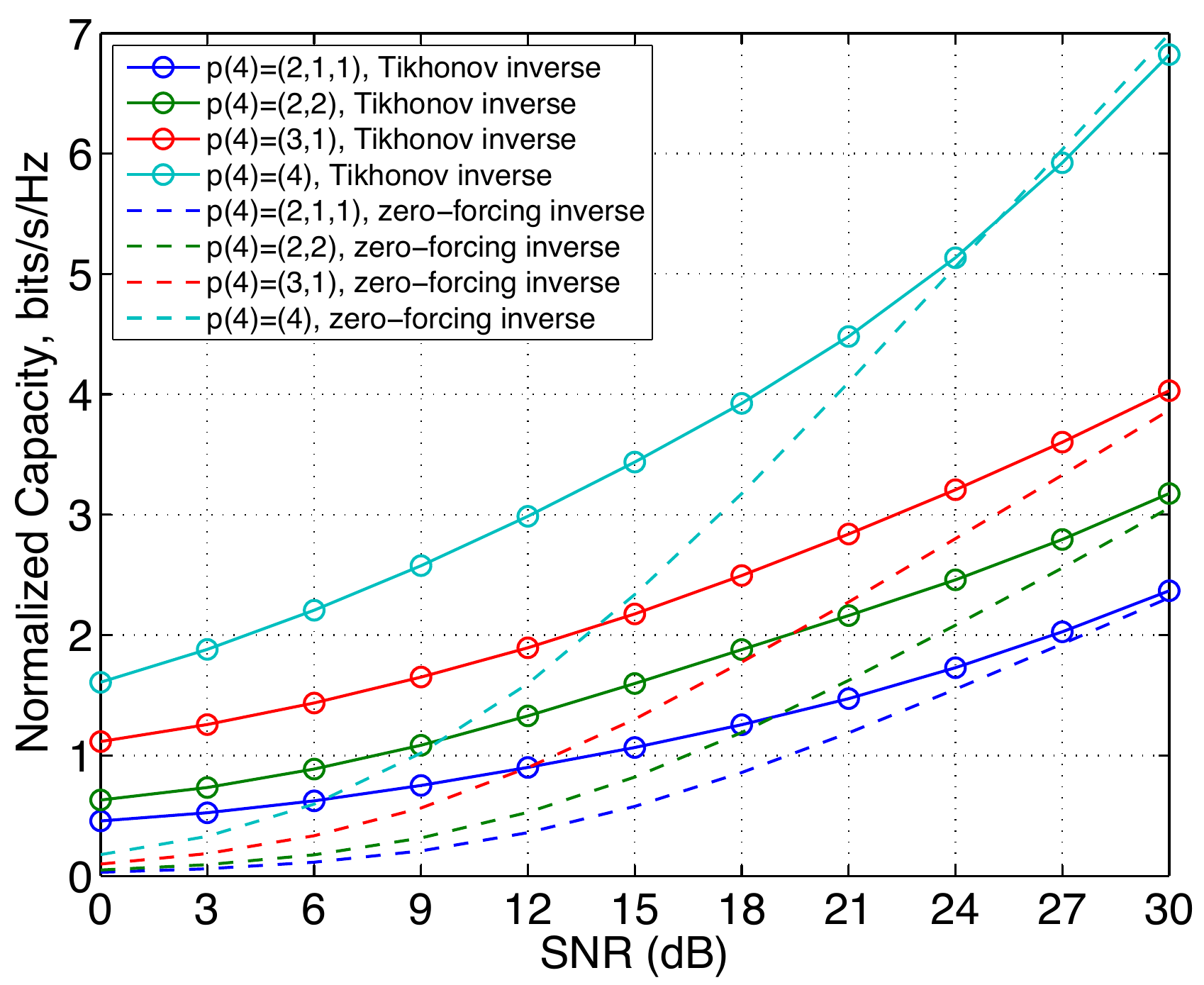}
\caption{Comparison of hybrid pre-inversion and interference alignment schemes for $K = 4$ users with $N = 5$ antennas for various partitions of the users, using the Tikhonov inverse with $\alpha = \sigma^2$ and the full subchannel inverses.  For every value of SNR, $10^4$ random Gaussian channel matrices were generated, and the average SINRs were calculated using (\ref{SINR_IA_def}).}
\end{figure}
This plot highlights the advantages of data sharing between users, as well as demonstrates the benefits of using the Tikhonov inverse in concert with an IA strategy as opposed to the zero-forcing inverse.  The apparent out-performance of the zero-forcing inverse when compared to the Tikhonov inverse for full channel pre-inversion at very high SNR is a numerical artifact of not having run enough trials.

\section{Conclusion}

We have proposed a hybrid interference alignment and channel pre-inversion scheme in which groups of user employ Tikhonov pre-inversion to clear interference from within their own group, and interference alignment to clear interference between groups.  For a single $N\times N$ MIMO channel employing Tikhonov inversion, we have provided an explicit formula which predicts the asymptotic behavior of the SINR as $N\rightarrow \infty$.  We used the case of $K = 4$ users each with $N = 5$ antennas as a testing ground, and completely classified the available degrees of freedom for every partition of the users into data-sharing groups.  Lastly, we have provided simulations which measure the ergodic capacities of the hybrid channel pre-inversion and interference alignment strategies we have constructed.  One obvious question which we leave for future work is designing joint interference alignment and channel pre-inversion strategies for different values of $K$ and $N$.

\section{Acknowledgements}

The first author is supported by Academy of Finland grant 268364.

\bibliographystyle{ieee}
\bibliography{myrefs_new.bib}

\end{document}